\documentclass[twocolumn,showpacs,preprintnumbers,amsmath,amssymb,aps]{revtex4}

\usepackage{graphicx}
\usepackage{dcolumn}
\usepackage{bm}

\newcommand{\binomial}[2]{\left(\begin{array}{c} #1\\ #2\end{array}\right)}

\begin{document}


\title{The phase transition in random catalytic sets}

\author{Rudolf Hanel$^{1}$}
\author{Stuart A. Kauffman$^{2,3}$}
\author{Stefan Thurner $^{4,}$}
\email{thurner@univie.ac.at}
\affiliation{
   $^{1}$ Institute of Physics; University of Antwerp; Groenenborgerlaan 171; 2020 Antwerp; Belgium\\
   $^{2}$ Biocomplexity and Informatics; The University of Calgary; Calgary; AB T2N 1N4; Canada\\
   $^{3}$Santa Fe Institute; 1399 Hyde Park; Santa Fe, NM 87501; USA \\
   $^{4}$ Complex Systems Research Group; HNO; 
   Medical University of Vienna; W\"ahringer G\"urtel 18-20; A-1090; Austria \\
   }
 
\begin{abstract}
The notion of (auto) catalytic networks has become a cornerstone in 
understanding the possibility of a sudden dramatic increase of 
diversity in biological evolution as well as in the evolution of social and 
economical systems.  
Here we study catalytic random networks with respect to the final 
outcome diversity of 
products. We show that an analytical treatment of this longstanding problem 
is possible by mapping the problem onto a set of non-linear recurrence equations. 
The solution of these equations show a crucial dependence of the final 
number of products on the initial number of products and the density 
of catalytic production rules. For a fixed density of rules
we can demonstrate the existence of a phase transition from a practically 
unpopulated regime to a fully populated and diverse one. 
The order parameter  is the number of final products. 
We are able to further understand the origin
of this phase transition as a crossover from one set of solutions 
from a quadratic equation to the other. 
\end{abstract}



\maketitle


\section{Introduction}
Chemicals act on chemicals to produce new chemicals, goods act on goods to 
produce new goods, ideas act on ideas to produce new ideas. The concept that 
elements of a set act on other elements of the same set to produce new 
elements which then become part of this set is ubiquitous not only in nature 
but also in social systems and processes. We might think of the 
development of modern chemistry, where the invention of a new compound 
leads to possibilities to use this compound (as a catalyst) to produce 
yet another compound. The same is true for economy, where one newly 
introduced good can be used as a tool to produce new goods and tools.
Other famous example are recent  models of 
evolution, maybe even our whole concept of history in general 
can be seen as a process of this type.

Maybe the most fascinating question associated with these processes is  
under which conditions self-sustaining systems can emerge, i.e. 
that the newly produced chemicals, goods, or ideas find adequate other new or old
chemicals, goods, and ideas such that they can act on each other to produce 
yet new products, and so on. For any scientific approach to this subject 
it is clear that it is necessary to specify rules, which product can act on 
another product to produce a third one. For example there is a chemical rule 
that oxygen and hydrogen will produce water but there is no rule that 
gold and helium can form a compound. There is a rule that a hammer acting on 
a block of iron will produce sheet metal, but no rule that welding 
together two blocks of uranium 238 will lead to a big block of uranium.
If one imagines that all existing and non-existing -- but possible -- 
products are listed in a high dimensional vector, than the rules how 
those elements can act on each other can be thought of elements in an 
interaction matrix, where a zero entry means no interaction, 
and a non-zero element gives the interaction strength. 

In history there have been plenty of instances where a system of the above type
underwent a transition from a state with very few products to a state 
of vast abundance of products. These transitions happen over relatively
short timescales. An example from biology is the Cambrian explosion 
\cite{cambrium}, where an unprecedented number of new taxa emerged within 
very short timescales. 
An economical example is the industrial revolution, where the number of 
industrial goods exploded to previously unimaginable numbers, 
a social example is the explosion of culture with the advent of the 
renaissance, or a more modern example the explosion of the number chemical 
compounds in the last century.  

All of these 'explosive' processes share the same structure: There are possibly 
constructively interacting elements, interaction is governed by a set 
of rules (natural laws, social consensus, religious restrictions), and 
a number of initially existing products. What does trigger the explosive 
event, why is this event sometimes missing for very long timeperiods? 
The only parameters at hand are: the number of rules, the number of 
initially present elements, and a possible structure in the interaction 
matrix.
It is likely that the cultural explosion was driven by an increase 
of rules which was possible by driving back religious restrictions. 
The explosion of chemical compounds was possible by discovering 
the rules of modern chemistry. 
Let us mention here that for well studied systems the set of rules can 
in principle be known completely. However, for large systems, it might 
be wise as a first step to model them stochastically, i.e. let the 
interaction matrix be a random matrix. In this case there will 
of course be no structure in the interaction matrix, and the only parameters 
will be rule density $r$ and initial number $a_0$ of existing elements.

Since quite some time there has been conjectured the 
existence of a phase transition in the above systems, 
e.g. in \cite{farmer1}.
By this we mean that in the $r-a_0$ plane there exist 
well defined regions,which are practically unpopulated, 
or almost fully populated, with a sharp transition between
the regimes. This means that fixing the number of initial 
elements there exists a critical density of rules $r_{crit}$. When 
the system is below  $r_{crit}$ the number of elements will 
remain relatively low compared to the total number of 
possible products. Above $r_{crit}$ the system will become 
self-sustaining and drive towards a heavyly populated state.

Even though this setting seems to be fundamental to a variety of 
disciplines, and its importance has long been noticed \cite{schumpeter11,order}, 
the progress of a systematic scientific treatment of these problems 
is limited. Relevant contributions to this field come from chemistry
and biology. 
The adequate mathematical treatment of such processes are so-called 
catalytic equations, which are sets of coupled, quadratic, ordinary 
differential equations. Special cases of those equations are 
for example the class of Lotka-Volterra replicators, the hypercycle \cite{eigen}, 
or more recent ideas like the Turing gas \cite{fontana}. 
Replicator dynamics, which is linear, is obtained from  catalytic 
dynamics by a proper scaling of time. 
 It is needless to mention that the non-linear catalytic equations 
carry tremendous potential of complex dynamics,  
however, in earlier studies some robustness in terms of 
fixed points seems to have been observed. For more details see 
\cite{stadler93}.  

The aim of the present work is to prove the existence and study  
details of the nature of the above mentioned phase transition. 
This is possible by introducing some concepts of set theory, and -- making use of the 
random structure of the production rules -- by mapping the size of 
a consecutive series of sets into a set of update equations, which 
can be solved and analyzed. The analytic formulas give insights into what 
is happening at the transition. A practical aspect of this present work
is that we cover analytically the large system limit, which is
beyond numerical reachability, and has sofar not been possible to 
study. 

The intuition for this work is based on a bit-string formulation 
of the above problem, which is a generalization of models recently 
termed random grammars \cite{order}. In the bit-string model there 
is a set of initial bit-strings and a set of strings which act on these 
strings by either combining strings or substituting sub-strings. 
These latter strings can be seen as catalysts, their existence 
constitutes the presence of rules of what can be combined and/or 
substituted. The model which is presented below can be shown to be one to one 
compatible with the substitution and combination rules in the 
Kauffman bit-string model.

The paper is structured as follows: Section 2, where the hard working authors
state the formal problem and develop the necessary notation and concepts. 
Section 3, where the brave authors derive the set of update equations.
Section 4, where the struggling authors succeed in solving these equations 
reasonably well, and Section 5, where the victorious authors present their   
results. 
In Section 6 we 
discuss the results.

\section{The network-equation and some notation}

Our basic concept is to view the abundances of all possible products 
as entries in a $d$-dimensional time-dependent vector $x(t)$, i.e. 
$x_i(t)$ is the quantity of product $i$ present at time $t$. We drop the 
time notation in the following. The population 
dynamics of a system, where product $i$ under the influence of product $j$ 
produces product $k$ is given by network-equations of the following type
\begin{equation}
\dot{x}_{k}=(x,\alpha^{k}x)-x_{k}\Phi \quad , \quad \Phi=\sum_{k}(x,\alpha^{k}x) \quad .
\label{cata}
\end{equation}
with $k\in\Lambda$, and the component $\alpha^k_{ij}>0$. 
$\Lambda$ is the domain of nodes -- or put in fancier terms -- 
the index set of the dynamic
process. $(,)$ is the inner product of statevector $x$ with 
dimensionality $d=|\Lambda|$, the number of all possible products or nodes.
The statevector is a vector of relative frequencies $0 \leq x_{k} < 1$, and
$\sum_k x_k=1$. 
Each $\alpha^{k}$ is a linear operator providing the quadratic 
forms $(.,\alpha^{k}.)$ and encodes the structural information about 
which binary combinations $(i,j)\in\Lambda^{2}$, the {\sl substrates}, can 
interact in order to form {\sl product} $k$. Each entry in $\alpha^k$ is
 a real number, however, for the sake of simplicity we consider 
only 1 for interaction or 0 for no interaction. $\alpha^k$ can be 
thought of as the set of rules how objects interact. In the following $\alpha^k$ 
is sampled as a random matrix in the following way: 
We assume that for each {\sl product} $k\in\Lambda$ there are
{\sl pairs of substrate} $(L(k),M(k))\in\Lambda^{2}$ such that 
$L(k)\stackrel{M(k)}{\longrightarrow}\{k\}$. In words, this arrow 
means: $k$ is produced by substrate $L(k)$ under the 'influence' of 
substrate $M(k)$. Note, that 
$L(k)$ and $M(k)$ do not have to be unique, there can be more than 
one pair $(L(k),M(k))$ producing a specific product $k$. Lets call the number 
of pairs leading to product $k$, ${\cal N}_{L,M}(k)$.
We define the {\it production rule density} as the average number of 
pairs leading to one product 
\begin{equation}
 r/2 = \langle {\cal N}_{L,M}(k) \rangle_k \quad.
\end{equation}

Equations like (\ref{cata}) are long known for their rather surprising robustness in terms of 
fixed points where the system converges to. This is not obvious and one would rather 
expect a situation more dominated by more complicated orbits and limit cycles. For a
more detailed discussion of this type of equations see \cite{stadler93}.
Fixed points will therefore provide much information about the
effective behavior of such systems, the fixed point equation being
\begin{equation}
x^{*}_{k}=\frac{1}{\Phi^{*}}(x^{*},\alpha^{k}x^{*}) \quad .
\end{equation} 
We are dealing with a nonlinear process, so that the solutions of the
network-equation will in general  depend strongly on the initial 
conditions. Even assuming the process to be driven towards a stable fixed point
does not imply the uniqueness of that fixed point.

\subsection{Notation}

We want to get some feeling what to expect when investigating the population dynamics
of randomly sampled fixed 0/1-networks. 
The dynamical properties of Eq. (\ref{cata}) are linked to specific topologies.
Knowing the  topologic features will enable one to solve for the dynamics.
Not knowing the interaction tensor $\alpha$ exactly, and  only given  
its statistical properties, following the classical concept of statistical physics,
one still can understand the expected dynamics of the system, and in 
particular its expected final outcome.
We hence study the system in a probabilistic fashion, i.e. 
averaging over all possible topologies. The choices for
$(L,M)(k)$ are equally distributed and $L$ and $M$ can be seen as independent
random variables 


The question we try to answer is: with how many products do we expect to
end up with when starting from a given number of randomly choosen 
initial substrate species?
Before we proceed we need a number of definitions.
We denote the number of elements contained in some set $A$ 
by writing $|A|$, and 
define the {\sl support} of a process at a given time 
\begin{equation}
S(x):= \left\{k| x_{k}>0 \right\} \quad .
\end{equation}
Suppose that the process is driven to a (stable) fixed 
point  such that the final support is
\begin{equation}
S^{*}(x):=S(\lim\limits_{t\rightarrow\infty}x(t)) \quad .
\end{equation}
Inversely, we can ask which initial conditions end up in the same fixed point
\footnote{If there is no stable fixed point the limits in the definition of the 
final support
are not converging and one has to specify convergent subsequences reflecting
the structure of the limit cycle or the chaotic regime.}
and define the body of this fixed point to be the set
\begin{equation}
B(x^{*}):=
\left\{x| \sum_{k}x_{k}=1 \wedge 
\lim\limits_{t\rightarrow\infty}x(t)=x^{*} \right\} \quad .
\end{equation}
We further define the following operations on product-sets:
the {\sl foreward difference} of set $A$,
\begin{equation}
\partial_{+} A:=\left\{k \in \Lambda| \{L(k),M(k)\}\subset A\right\}\setminus A
\end{equation}
 and the {\sl backward difference} of $A$,
\begin{equation}
\partial_{-} A:=\cup_{k\in A}\{L(k),M(k)\}\setminus A  
\quad .
\end{equation}
With these operations we can define the {\sl foreward closure} of set $A$
\begin{equation}
\bar{A}^{+}:=\cap\left\{B|A\subset B\wedge \partial_{+} B\subset B\right\} \quad ,
\end{equation}
and the {\sl backward closure} of a set,
\begin{equation}
\bar{A}^{-}:=\cap\left\{B|A\subset B\wedge \partial_{-} B\subset B\right\} \quad .
\end{equation}
These definitions can best be understood by viewing our dynamical system Eq. (\ref{cata}) as 
a directed graph whose nodes are connected by arrows. A node looking at 
its edges can therefore distinguish between feather ends or arrow heads. A node holding a feather end 
of an arrow is a substrate, a node holding an arrow head is a product with respect to this edge. 
Given some set of nodes  we can identify for each node $k$ all the nodes
which will be formed due to the influence of $k$ in the next timestep.
In this sense we say the nodes look 
'foreward'.
The foreward difference of some set is therefore the set of all 
products the nodes of the set look foreward to, excluded the products that 
are already present as nodes of the set we started out with. 
The backward difference follows the same idea 
only looking at the arrows from arrowhead towards the feathers.
The corresponding closures are then simply obtained by adding the set-differences 
iteratively to the initial set, i.e.
we add the difference to the initial set and form a new difference on the 
first extension of the set, then add this difference to obtain the 
second extension of the initial set and so on, until the
difference (forward or backward) is empty and the iterative process 
comes to a halt.
That this eventually may happen can be understood by looking at chains of arrows
as an example. Take a node and add an arrow pointing at some other randomly 
chosen node in the set. In the beginning the chance to select a virginal 
node is high and the chain will grow
but as the chain grows the probability of sampling a member already in the chain 
increases, and even though the single sample probability may still 
be small, the chance to sample into the chain eventually is not. 

For clarity let us summarize our philosophy: 
All objects which can possibly act  on each other are
represented by the index set $\Lambda$ of the processing system Eq. (\ref{cata}), which 
contains all the 'names' of the considered and 'thinkable' objects. The index set provides the
domain for all dynamical considerations. Properties of the system are implemented via
the map $x:\Lambda\rightarrow{\bf R}_{+}$, which
are the relative frequencies of the indexed species in the domain. We are
not interested in particular weights but only in the directed network 
topology coded by the matrices $\alpha^{k}$
on the domain $\Lambda$. Not only does this 
topological approach provide us with the means to talk about randomly
distributed productive substrate pairs, but also about their density of
occurrence $r$. 
Even more important, if we are not interested in details of the dynamics 
but decide to focus only on how 
large the expected final set will be (given some initial substrate set and density)
we can drop  dynamical considerations
and pass to topological ones.  We are not interested in how much of each object species
we will effectively end up with, just if it got produced or not. 
The subset of the domain that is
effectively populated is called the support $S\subset\Lambda$. To
investigate the flow of the support under the network equation we may 
utilize set-operations compatible with the topological structure of directed
reaction graphs leading to the definition of the forward and the backward
difference and their respective closures. It is intuitively clear, that the
forward closure of some set is the subset of the domain that is flooded by the
initial set during the production process. It necessarily forms an upper limit 
for the size of possible self-sustainable subsets of the domain reachable from 
the initial condition. In a simplification considering an equally 
distributed random interaction 
topology we gain a notion of expected  growth rates for the set-differences 
based on expected sampling rates. This leads to equations of expected growth
as demonstrated next.

\section{The growth laws for catalytic sets}

We now develop a method to compute the size of the foreward
closure $a_{\infty}=|\bar A^+|$ as a function of the production rule density and the 
size of the initial set $A$. $a_{\infty}$ is the final number 
of products once the system has converged.
The probability of some fixed $k$ being the product of some fixed pair of substrates, 
$l$ and $m$ is obviously 
\begin{equation}
   p=\frac{2}{d(d-1)} \quad .
\end{equation}
\begin{figure}[tbh]
\begin{tabular}{c}
\includegraphics[width=8.0cm]{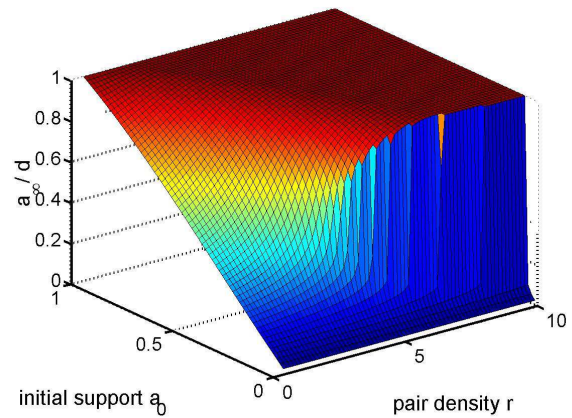} \\
{\Large (a)}\\
\includegraphics[width=8.0cm]{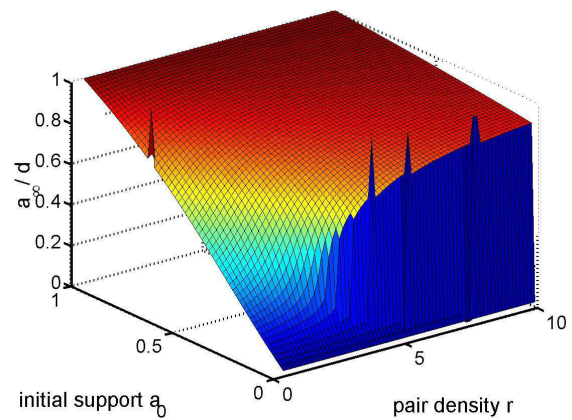} \\ 
{\Large (b)}\\
\end{tabular}
\caption{ Solution to the update equations Eq. (\ref{growth}) (a) and its analytical 
approximation Eq. (\ref{res}) (b). There is clearly a critical line in the
$r-a_0$  plane where a transition from an unpopulated to an almost fully populated 
regime occurs.  The spikes  in (b) are due to numerical problems 
in the plot with poles in the solution $a_{\infty}$. }
\label{Fig_fw}
\end{figure} 
Imagine an initial random set $A_0$ of products with a number of
 $a_0=|A_0|$ elements.
Given the dimensionality $d$ of the system and a production rule 
density $r$, the expected number of produced elements in the next timestep 
is the number of possible pairs in $A_0$ times the probability to find 
a productive pair, i.e.
\begin{equation}
\frac{rd}{2}\binomial{a}{2} p =\frac{ra(a-1)}{2(d-1)} 
	\sim\frac{ra^{2}}{2d} \quad .
	\label{propair}
\end{equation} 
Several of these newly created elements will already be $\in A_0$. 
The probability that one of these produced elements is not yet $\in A_0$
is $1-a_0/d$, leading to the actual size of the catalytic set at time 
1, $A_1$
\begin{equation}
  a_1=a_0+\Delta a_0 \quad {\rm with} \quad \Delta a_0= 
  \frac{r}{2d}\left(1-\frac{a_{0}}{d}\right) a_{0}^{2} \quad ,
\end{equation}
with $\Delta a_0 = |\partial_{+}A_{0}|$ being the increment of elements. 
What will happen in the next timestep? We now have 
$A_{1}=A_{0}\cup\partial_{+}A_{0}$. The increment for the 
next timestep will be made up of all the new products which are 
possible (and not already in $A_1$) by combining two elements 
from $\partial_{+}A_{0}$, or 
by  combining one element from $\partial_{+}A_{0}$ with one from 
$A_0$. The number of possible pairs for those combinations are
$\Delta a_0(\Delta a_0-1)/2\sim\Delta a_0^2/2 $ and $a_0 \Delta a_0$, respectively.
Note, that the third possibility combining 2 elements from the set 
$A_0$ will always lead to $\partial_{+}A_{0}$  which is already $\in A_1$, 
and no new products can emerge from this.
These possibilities multiplied by $p$, $rd/2$ and probability that the 
new product lies in $A_1$ already, lead to the increment for the second timestep
\begin{eqnarray}
  \Delta a_1= 
  \frac{r}{2d}\left(1-\frac{a_{1}}{d}\right)
  \left( \Delta a_{0}^{2} + 2 a_0 \Delta a_0 \right) \nonumber \\
  =\frac{r}{2d}\left(1-\frac{a_{1}}{d}\right)
  \left(  a_1^2 - a_0^2 \right) 
  \quad .
\end{eqnarray} 
Now continue the iterative sceme $A_{t+1}=A_{t}\cup\partial_{+}A_{t}$.
This means that the set at time $t+1$ is the old set plus the 
products newly generated in the timespan $[t,t+1]$.
We can finally write down a growth equation for catalytic sets 
\begin{equation}
 a_{t+1}=a_{t}+\Delta a_{t} \,\, , \,\,
 \Delta a_{t+1}=\frac{r}{2d}\left(1-\frac{a_{t+1}}{d}\right)
 \left(a_{t+1}^{2}-a_{t}^{2}\right)  \,\, ,
\label{growth}
\end{equation} 
with initial conditions $a_{0}=|A_0|$ and $a_{-1}=0$. 
To sum up, we have to take care of all possible new pairs by
looking at all new elements added in an  iteration step and 
the new pairs they can build with themselves. We also  have to take
into account all the pairs they can build with elements produced earlier.
We have to exclude all the pairs that allready have 
been considered. This is all captured by Eq. (\ref{growth}), which should be noted to be 
scale invariant with respect to dimension $d$.
To see this just scale $a \longrightarrow a/d$, and the dimension drops out of the 
equation. It is  therefore fully justified to drop $d$ from Eq. (\ref{growth}), if wanted.

\section{Solution of the update equation - Result}

The solution of the growth equation Eq. (\ref{growth}) 
with respect to the productive pair density $r$ and the 
initial set-size $a_{0}$ is given in Fig. \ref{Fig_fw} (a).
The immediate message is that there is a critical density $r_{crit}$ above 
which a continuous increase of the initial set-size $a_0$ does not 
correspond with a continuous increase of its foreward closure 
size $a_{\infty}$, but displays a phase transition, a jump from small to very
large foreward closures at some $a_0^{crit}(r)$. 
$a_0^{crit}$ vanishes for $r<r_{crit}$.
This demonstrates unambiguously the existence of a phase transition
in catalytic random systems. 

\subsection{Analytical approximation of the foreward closure size }

\begin{figure}[htb]
\begin{tabular}{c}
\includegraphics[width=7.0cm]{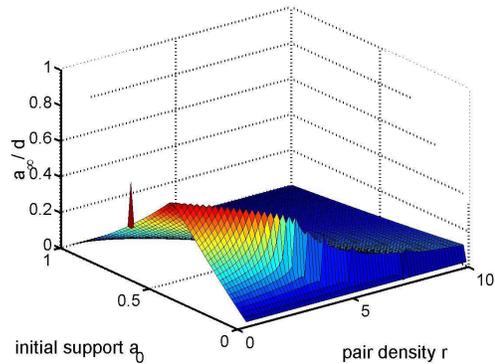} \\
{\Large (a)}\\
\includegraphics[width=7.0cm]{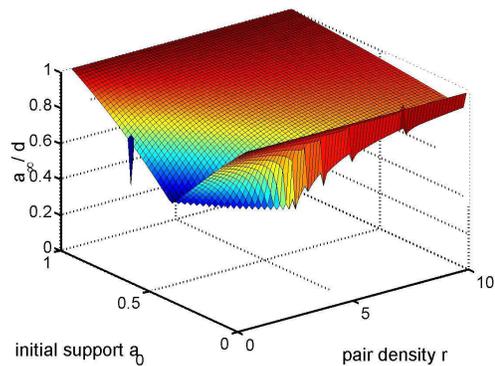} \\ 
{\Large (b)}\\
\includegraphics[width=7.0cm]{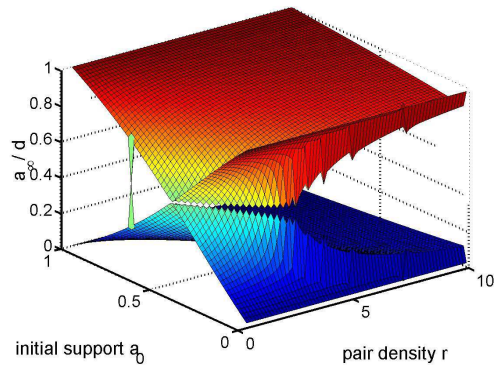} \\ 
{\Large (c)}\\
\end{tabular}
\caption{Plot of the solutions $a_{\infty -}$ (a),  $a_{\infty +}$ (b) and both of the 
above in one plot (c). It is clear that the pase transition happens as 
the solutions switch from $a_{\infty +}$ to $a_{\infty -}$.}
\label{Fig_pm}
\end{figure} 
 
Let us define  $c_{t}=\Delta a_{t+1}/ \Delta a_{t}$. 
As an approximation imagine for a moment that $c_t$ 
is a constant in time $c$. Then for the forward closure  
\begin{equation}
 a_{\infty}= \sum_t c^t a_0 = \frac{1}{1-c} \quad .
\end{equation}
A choice $c$ which produces the right asymptotical results is 
\begin{equation}
 c=r\left(1-\frac{a_{\infty}}{d}\right)\frac{a_{\infty}}{d} \quad ,
\end{equation}
which is seen by first dividing the right part of Eq. (\ref{growth}) by $\Delta a_t$ 
and then taking the $t\longrightarrow \infty$ limit.
Now  use the  Ansatz
\begin{equation}
 a_{\infty}=\frac{a}{1-c} \quad , \quad 
c=r\left(1-\frac{a_{\infty}}{d}\right)\frac{a_{\infty}}{d} \quad ,
\end{equation} 
which leads to a polynomial of third degree 
$a_{\infty}^{3}-a_{\infty}^{2}d+a_{\infty}\frac{d^{2}}{r}
-\lambda a\frac{d^{2}}{r}=0$, 
or in terms of $c$
\begin{equation}
c^{3}-2c^{2}+c(1+r\frac{a}{d})+r\frac{a}{d}\left(\frac{a}{d}-1\right)=0 \quad .
\end{equation}   
Substituting $c=x+2/3$ reduces to equation $x^{3}+px+q=0$ with 
\begin{equation}
p=-\frac{1}{3}\left(1-3r\frac{a}{d}\right)\qquad
q=\frac{2}{27}+\frac{1}{3}r\frac{a}{d}\left(3\frac{a}{d}-1\right) \quad ,
\end{equation}   
so that  Cardano's method can be used with $x=y+z$ to get 
\begin{equation}
y=\left[ -\frac{q}{2}+\sqrt{(q/2)^{2}+(p/3)^{3}} \right]^{\frac13}\qquad z=-\frac{p}{3y} \quad ,
\end{equation} 
providing us with three solutions $y+z$, $\rho^{2}y+\rho z$ and 
$\rho y+\rho^{2} z$ with $\rho=\exp(2\pi i/3)$.  We are only 
interested in the real solution $y+z$ 
which yields the final result
\begin{equation}
a_{\infty}=\frac{a}{1-c} \quad {\rm with} \quad c=2/3+y+z \quad ,
\label{res}
\end{equation}
which is a good approximation of the true solution of the forward equation 
Eq. (\ref{growth}) and  is plotted in Fig. \ref{Fig_fw} (b) 
in the $a_0$--$r$ plane. The comparison with Fig. 1 (a) demonstrates
the quality of the analytical solution. 

The analytical form of the solution allows us to understand the origin of 
the phase transition. For this reason we observe that we  
can compute $a_{\infty}$ by solving $c=r(1-a_{\infty})a_{\infty}$
providing us with two solutions
\begin{equation}
a_{\infty \, \pm}=\frac{d}{2}\left(1\pm\sqrt{1-4\frac{c}{r}}\right) \quad .
\end{equation}   
These solutions are shown in Figs. \ref{Fig_pm} (a) and (b).
It becomes obvious that the  phase transition actually always takes 
place by the system switching from the one solution to the other at 
$a_0^{crit}(r)$. Below  a critical value of $r_{crit}$ the one solution 
smoothely passes into the other. We observe a crossover of the 
probable and the {\sl improbable} solution.
Most likely this is due to a convexity argument, implying the 
monotonicity of the size of the forward closure with respect to 
the size of the initial condition. The size of the forward closure 
will not shrink with increasing the initial set size
and therefore has to jump to the alternative solution 
at some critical line.
As long as the third order polynomial is strictly
monotonic we have a unique real solution for the zero of the polynomial.
At $r_{crit}$ the polynomial starts to have a local 
minimum and maximum and we have in fact two relevant real zeros (out of three),
the large and the small solution. $r_{crit}$ is determined by the 
tripple zero of the polynomial. 
Note a famous analogy here: Just as in the Vanderwaal's gas we can draw
Maxwell lines. At the phase-transition, the small solution becomes
instable, the large solution becomes stable. Here  the
productive pair density $r$ takes the role of temperature in the Vanderwaal's gas. Below the critical
value a system {\sl freezes} into a small set of durable species, 
while above the critical production law density, supercritical but yet small sets of 
individuals {\sl evaporate} into their foreward closure. 
The difference we have here compared to Vanderwaal's gas is that the 
{\sl liquid} phase is concentrated on the high edge of the phase 
transition, while the phase-plane itself is a solid-liquid mixture.

\section{Discussion and outlook}

We have developed a way to map a conceptual system 
of autocatalytic agents into a quantitative framework.
With this it is possible to show that the combination of 
the initial number 
of products $a_0$ and the density of mutual production rules $r$ is 
crucially influencing the mode of growth of sets of products. 
Our main result is that we were able to map the class 
of random catalytic networks into a set of
growth equations, which allows us to study the expectation 
value of the final number of products. 
The resulting equation can be solved and shows a 
clear phase transition from practically unpopulated zones towords 
almost fully populated ones, in 
'rule density--initial number of products' space. 
The transition is a crossover between two sets of 
solutions to a quadratic equation.

We believe  that this nicely resembles numerical-experimental findings of 
\cite{stadler93}, where the authors  find a decrease of species 
with an increase of $p_{el}$ in their Fig. 1 (d). The fact 
that the transition from small to full forward closure sizes 
is gradual, is either due to the limited system size (10 species), 
or that the initial support was high, i.e. above $a_0^{crit}(r)$,
so that no sharp increase could be found.
In fact, systems of the type of Eq. (\ref{cata}) are straight forwardly 
solvable numerically up to system sizes of about $d=100$ within 
reasonable computing time.
We note that the practical value of this present work is that
it captures the large system size limit, which is out of numerical reach 
and was not tractable analytically before.

The present approach does not try to explain the detailed role 
of auto catalytic cycles nor of keynode species in the context 
of understanding the beginning of an explosion of species 
numbers as was very nicely done for example in \cite{jain99,jain02}. 
The periods of fast extinction reported there are not incorporated 
in our consideration yet, since we have not taken any evolutionatry 
hypotheses into account so far, nor did we incorporate evolutionary concepts in the sense 
that products are associated characteristics (some sort of fitness, 
e.g. the weight $x_k$) upon which they can get selected by some 
method. Taking these arguments into account seems a natural starting 
point for future research. 
We believe that it should be possible that 
a combination of backward and forward closure arguments can be 
used to estimate a critical density of auta catalytic cycles necessary 
for a system to become critical as studied here. 
We have not so-far considered negative entries in the interaction 
matrices, which should also be present in natural or social systems.
Finally, we mention that our arguments given here should -- due to 
the absence of a characteristic scale in the update equations -- 
not be limited to finite sets. \\

\noindent
{\bf Acknowledgments:}
S.T. would like to thank the SFI and in 
particular J.D. Farmer for their great hospitality and support 
in Sept-Oct of 2004. The project was funded in part by the Austrian 
Council, and FWF project P17621 G05.  



\end{document}